\documentclass[sigconf]{acmart}
\AtBeginDocument{%
  \providecommand\BibTeX{{%
    \normalfont B\kern-0.5em{\scshape i\kern-0.25em b}\kern-0.8em\TeX}}}
    
\usepackage{booktabs} % For formal tables
\usepackage{setspace}
\usepackage{enumitem}
\usepackage{microtype}
\newcommand{\tabincell}[2]{\begin{tabular}{@{}#1@{}}#2\end{tabular}}
\usepackage{algorithm}
\usepackage{multirow}
\usepackage{tabularx}
\usepackage[utf8]{inputenc} 
\usepackage{multirow}
\usepackage{float} 
\usepackage{amsmath}
\usepackage{graphicx}
\usepackage{subfigure}
\usepackage{epstopdf} 
\usepackage{makecell}
\usepackage{listings}
\usepackage{tablefootnote}
\usepackage[flushleft]{threeparttable}
\usepackage{diagbox}
\usepackage{booktabs}
\usepackage{siunitx}
\usepackage{footmisc}
\usepackage{footnote}
\usepackage{enumitem}
\usepackage{tabularx}
\usepackage{graphicx}  % 可能需要用到的包
\usepackage{booktabs}  % 让表格更美观
\usepackage{makecell}  % 允许单元格内换行
\usepackage{multirow}  % 可选：支持跨行单元格
\usepackage{lscape}    % 可选：支持横向表格
\usepackage{makecell}
\usepackage[normalem]{ulem}
\usepackage{colortbl}
\usepackage[utf8]{inputenc}
\usepackage{threeparttable}
\usepackage{CJKutf8}
\usepackage[utf8]{inputenc}
\usepackage{makecell}
\usepackage{algpseudocode}
\useunder{\uline}{\ul}{}

\definecolor{red1}{rgb}{1,0.5,0.5}
\definecolor{red2}{rgb}{1,0.7,0.7}
\definecolor{red3}{rgb}{1,0.8,0.8}
\definecolor{red4}{rgb}{1,0.9,0.9}
\definecolor{red5}{rgb}{1,0.95,0.95}

\definecolor{b0}{rgb}{0.1,0.2,1} % 1
\definecolor{b1}{rgb}{0.3,0.5,1} % 2
\definecolor{b2}{rgb}{0.5,0.7,1} % 3
\definecolor{b3}{rgb}{0.6,0.8,1} % 4
\definecolor{b4}{rgb}{0.85,0.95,1} %

\definecolor{g0}{rgb}{0.1,1,0.2} % 1
\definecolor{g1}{rgb}{0.3,1,0.5} % 2
\definecolor{g2}{rgb}{0.5,1,0.7} % 3
\definecolor{g3}{rgb}{0.88,1,0.88} % 4
\definecolor{g4}{rgb}{0.85,1,0.95} %
\definecolor{mygray}{gray}{.9}

\usepackage{lipsum}
\begin{document}

% \begin{CJK}{UTF8}{gbsn}

\title{Qilin: A Multimodal Information Retrieval Dataset with APP-level User Sessions}

\author{Jia Chen$^\dag$}
\affiliation{%
  \institution{Xiaohongshu Inc.}
}
\email{chenjia2@xiaohongshu.com}

\author{Qian Dong$^\dag$}
\affiliation{%
  \institution{Tsinghua University}
%  \institution{Xiaohongshu Inc.}
}
\email{dq22@mails.tsinghua.edu.cn}

\author{Haitao Li}
\affiliation{%
  \institution{Tsinghua University}
%  \institution{Xiaohongshu Inc.}
}
\email{liht22@mails.tsinghua.edu.cn}

\author{Xiaohui He}
\affiliation{%
  \institution{Xiaohongshu Inc.}
}
\email{manyou@xiaohongshu.com}

\author{Yan Gao}
\affiliation{%
  \institution{Xiaohongshu Inc.}
}
\email{yadun@xiaohongshu.com}

\author{Shaosheng Cao}
\affiliation{%
  \institution{Xiaohongshu Inc.}
}
\email{shelsoncao@gmail.com}

\author{Yi Wu}
\affiliation{%
  \institution{Xiaohongshu Inc.}
}
\email{xiaohui@xiaohongshu.com}

\author{Ping Yang}
\affiliation{%
  \institution{Xiaohongshu Inc.}
}
\email{jiadi@xiaohongshu.com}

\author{Chen Xu}
\affiliation{%
  \institution{Xiaohongshu Inc.}
}
\email{chenlin1@xiaohongshu.com}

\author{Yao Hu}
\affiliation{%
  \institution{Xiaohongshu Inc.}
}
\email{yaoohu@gmail.com}

\author{Qingyao Ai}
\affiliation{%
  \institution{Tsinghua University}
}
\email{aiqy@tsinghua.edu.cn}

\author{Yiqun Liu}
\affiliation{%
  \institution{Tsinghua University}
}
\email{yiqunliu@tsinghua.edu.cn}
\renewcommand{\shortauthors}{J. Chen et al.}

% 密集的交互性。由于用户往往缺乏基本的法律知识，法律咨询通常需要多轮互动，由从业者逐步指导用户。

\begin{abstract}

User-generated content (UGC) communities, especially those featuring multimodal content, improve user experiences by integrating visual and textual information into results (or items).
The challenge of improving user experiences in complex systems with search and recommendation (S\&R) services has drawn significant attention from both academia and industry these years.
However, the lack of high-quality datasets has limited the research progress on multimodal S\&R.
To address the growing need for developing better S\&R services, we present a novel multimodal information retrieval dataset in this paper, namely \textsf{Qilin}.
The dataset is collected from \textit{Xiaohongshu}, a popular social platform with over 300 million monthly active users and an average search penetration rate of over 70\%. 
In contrast to existing datasets, \textsf{Qilin} offers a comprehensive collection of user sessions with heterogeneous results like image-text notes, video notes, commercial notes, and direct answers, facilitating the development of advanced multimodal neural retrieval models across diverse task settings. 
To better model user satisfaction and support the analysis of heterogeneous user behaviors, we also collect extensive APP-level contextual signals and genuine user feedback. 
Notably, \textsf{Qilin} contains user-favored answers and their referred results for search requests triggering the Deep Query Answering (DQA) module.
This allows not only the training \& evaluation of a Retrieval-augmented Generation (RAG) pipeline, but also the exploration of how such a module would affect users' search behavior.
Through comprehensive analysis and experiments, we provide interesting findings and insights for further improving S\&R systems.
We hope that \textsf{Qilin} will significantly contribute to the advancement of multimodal content platforms with S\&R services in the future.

\end{abstract}
\maketitle
\section{Introduction}

% \begin{CJK}{UTF8}{gbsn}
% 中文
% \end{CJK}
%\footnotetext[1]{These authors contributed equally to this work.}

% 第一段，写搜索和推荐系统是社交平台重要的组成部分，然而大多数已有的训练数据或者基准集合都是文本模态的，不能很好地支持多模态搜推系统的训练和评测。
Search engines and recommender systems play an essential role in online content platforms nowadays.
Beyond textual content, mainstream User-Generated Content (UGC) communities usually provide illustrated texts or videos as results in either single-column or two-column display~\cite{zhang2024notellm,gao2022kuairec}.
These multimodal contents help users find desired information more conveniently, e.g., a note with images for each step of preparing a Tiramisu or even a video for the whole process is more intuitive for users to reproduce the flavor.
Therefore, how to incorporate multimodal elements into the retriever is crucial for improving system effectiveness.
Nevertheless, most existing datasets for general search or recommendation tasks mainly contain textual information or statistically dense features, which is deficient for investigating better multimodal search and recommendation (\textbf{S\&R}) services.

\begin{figure}[t]
\centering
\includegraphics[width=\linewidth]{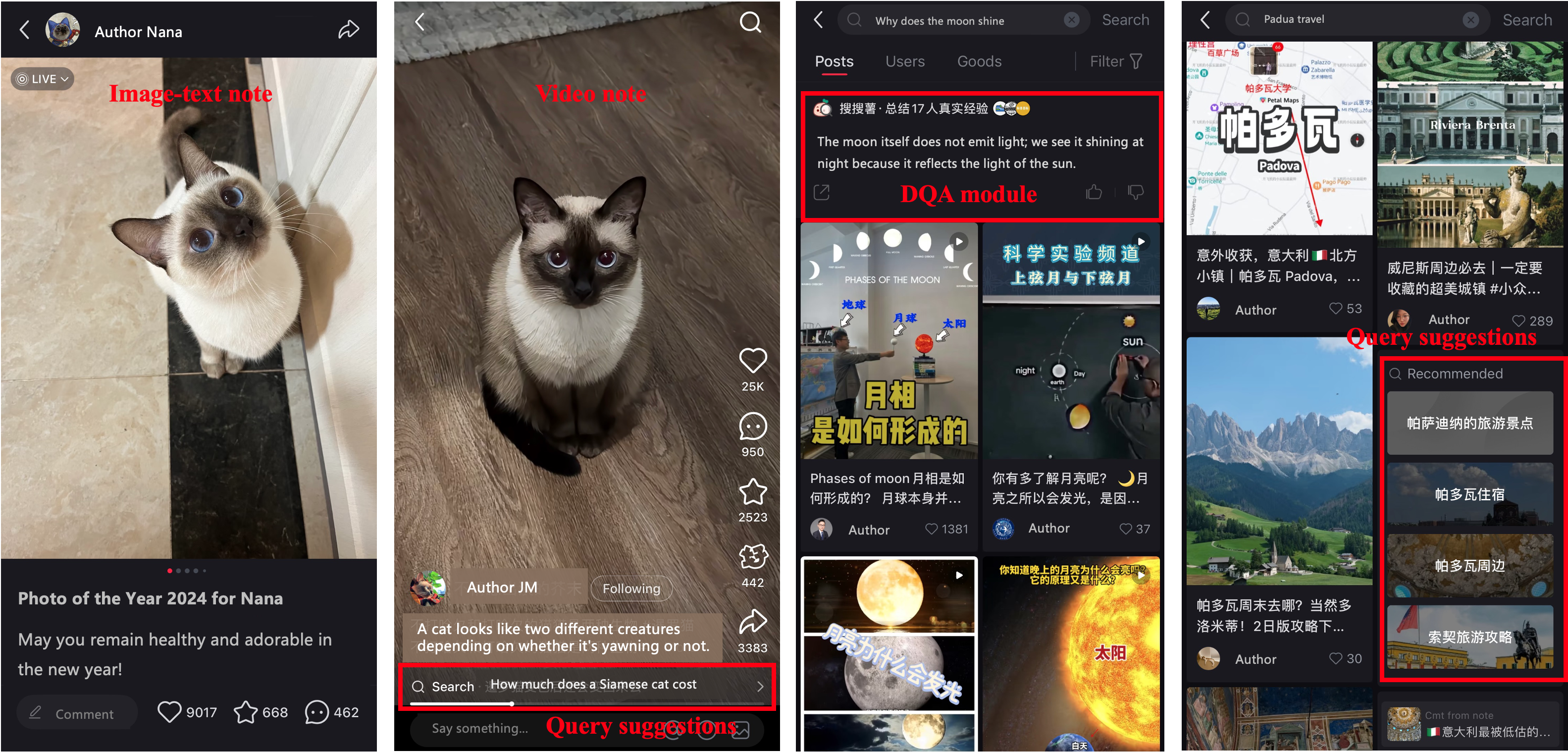}
\vspace{-0.2cm}
\caption{Xiaohongshu leverages a two-column result list for S\&R services, retrieving heterogeneous results like image-text, video, and commercial notes. The search system is equipped with a DQA module to provide direct answers for users. There are also various modules to stimulate users to search for any topics they might be interested in.}
\label{app_tracks}
\end{figure}

% 第二段，已有的很多研究更多地专注于单次请求下的系统性能，对用户的中长期建模不够，不能够充分优化用户的满意度乃至长期的留存。我们需要更多地关注session上下文信息、用户搜索意图来源、用户状态转移模式、用户回访行为等。例如，我们很难直接建模用户的长期留存，但在经验性分析中，用户回访行为和长期留存高度相关。另外，除了点击、点赞等消费指标外，用户的换词行为也是满意度的重要代理信号。因此，提供一份app-level的用户行为数据，对于分析用户满意度和留存是非常重要的。（未来也可以通过三方标注把该数据集补充为一个评价指标研究数据集）
% 看看第一页这个图要怎么画‼️‼️

%To improve user satisfaction for a specific mobile application, a normal paradigm is to separately optimize each module or service.
To improve user satisfaction for a specific mobile application, in-depth user behavior analysis at the APP level can be crucial.
As users' information needs become complicated, they may issue a series of related queries within a short interval to strive for better results, a.k.a., session search~\cite{ahmad2019context,chen2019tiangong,chen2020context}.
In other circumstances, users may be inspired by a recommended result or a query suggestion module~\cite{chen2021towards,ozertem2012learning} to explore relevant topics via the search system.
Users' APP-level tracks are very diverse, containing abundant contextual information.
Some of these contexts/factors can be essential for estimating user satisfaction or long-term retention, e.g., ~\citet{ozertem2012learning} find that besides click signals, users' reformulating behavior can be regarded as a good proxy for modeling satisfaction.
Therefore, beyond a single request, we should also focus on the session-level signals such as reformulating \& revisiting actions, the source of user search intent, and the pattern of APP-level user transition behaviors (\textbf{S}$\rightarrow$\textbf{R} or \textbf{R}$\rightarrow$\textbf{S}) to better estimate long-term user satisfaction towards the whole application.

% 第三段，大模型时代的到来，让搜索&推荐系统有了更多可能。目前大部分的搜索系统会配上一个DQA模块给用户提供一个直接的回答，提高用户的搜索效率。然而，此类模块对用户满意度和留存的影响仍然是未知的。例如，用户在点击DQA之后，对自然结果的点击变少了，如何更好地评价相关RAG模块对整体系统的影响，以及正确评估RAG系统的性能，需要有对应的数据来支撑。
Due to the different forms of information displayed in multimodal scenarios, traditional behavioral analysis results may no longer be applicable.
Predicting user satisfaction in a multimodal S\&R system with heterogeneous functional modules is still challenging.
For example, modern search engines usually incorporate a Deep Query Answering (DQA) module to provide users with succinct, direct answers.
Typically, a DQA module operates through a Retrieval-Augmented Generation (RAG) pipeline~\cite{gao2023retrieval,wang2024searching} which firstly retrieves relevant documents and then prompts a LLM for self-consistency check and summarization.
When a direct answer is presented, user browsing behavior will be greatly impacted, e.g., they may focus more on the top results while interacting less with other organic ones~\cite{wu2020providing}.
De facto, the influence of the DQA module on user satisfaction and retention still remains largely under-investigated.
To better evaluate the effect of the RAG module on users' perceived experience, the academia highlights the need for a dataset that includes both genuine user feedback on the DQA module and contextual user behaviors before and after they engage with such a module.

% 第四段，为了能更好的促进新一代搜推系统的发展，我们提出了一个创新的多模态数据集Qilin。主要的贡献有1、2、3，，，
To shed light on the aforementioned issues, we present a novel multimodal dataset with large-scale APP-level user information discovery sessions in multiple scenarios (including search, DQA, and recommendation), namely \textsf{Qilin}.
All user sessions are collected from \textit{Xiaohongshu}~\footnote{Known in English as \textit{rednote}, official site: \url{www.xiaohongshu.com}.}, which is a popular social platform as well as the largest lifestyle search engine in China, with over 300 million monthly active users and a search penetration rate of over 70\%.
Different from other platforms, \textit{Xiaohongshu} provides heterogeneous results such as image-text notes, video notes, commercial notes, and direct answers (as shown in Figure~\ref{app_tracks}), posing challenges for optimizing both search and recommendation.
%这里要加一些数据集的细节，等数据schema完成之后加下
To sum up, the main novelties of \textsf{Qilin} are listed as follows:
\begin{itemize}[leftmargin=*]
	\item To the best of our knowledge, \textsf{Qilin} is the first practical multimodal S\&R dataset with heterogeneous results collected from a social media platform. Genuine user behaviors and multimodal content features facilitate the training of sophisticated neural architectures and even LLMs in various task settings. 
	\item Besides side information like request and item features, we also contain abundant APP-level contextual signals (e.g., query sources, request history, timestamps, position, etc) and multiple user feedback toward the whole system. These signals are crucial for the in-depth investigation of user state transitions, revisits, and query reformulations to model user satisfaction or long-term retention. 
	\item For the search requests triggering the DQA module, we further collect user-favored answers and their referred results as the positive instance under the specific query. To this end, \textsf{Qilin} can not only be taken as a benchmark for training or evaluating an RAG module but also be used for exploring the influence the module brings to user behavior.
\end{itemize}

To facilitate the reproducibility of this work, all resources for \textsf{Qilin}, code implementation, and related experiment details have been released in the repository below~\footnote{\url{https://github.com/RED-Search/Qilin}}.

\section{Related Work}

\subsection{Multimodal Information Retrieval}
Multimodal information retrieval has been extensively studied by Information Retrieval (IR), Computer Vision (CV), and Multimedia communities.
To facilitate understanding, we systematically categorize related studies by approaches and task types, respectively.
% 这块太复杂了，可以画个图

Generally, multimodal retrieval approaches can be divided into three groups: 1) \textit{representation learning} based approaches~\cite{zhu2020deep,frome2013devise,balaneshin2018deep,wang2016effective,wei2021universal,laenen2018web}, 2) \textit{modality fusion or interaction} based approaches~\cite{liu2018attentive,xie2019improving,guo2018multi}, and 3) \textit{hybrid modeling} approaches~\cite{qu2021dynamic,liu2023multimodal,guo2018multi,liu2021que2search}.
Representation learning-based approaches aim to map a modality (usually an image) into a binary Hamming space with hash functions~\cite{luo2018scalable,zhu2020deep} or to encode it into a latent semantic space~\cite{frome2013devise,balaneshin2018deep}.
Generally, hash-aware methods have low storage costs and are efficient in real-time infrastructure~\cite{yang2017visual,zhang2018visual}.
In contrast, semantic-based ones focus more on modality understanding and cross-modality matching with deep neural networks (DNNs).
Although DNNs show better generalization compared with traditional approaches~\cite{zhang2013attribute}, architectures with modality fusion and interaction usually achieve higher performance.
%For instance, ~\citet{guo2018multi} map textual and visual user preferences into a common latent space with feature fusion and comparative learning.
To address intra-modal reasoning and cross-modal alignment, ~\citet{qu2021dynamic} develop a dynamic modality interaction network with a routing mechanism.
Consequently, the hybrid modeling of combining modality representation with multimodal feature interaction tends to outperform single modeling approaches.

For task type, related literature can be classified into \textit{cross-modal retrieval}~\cite{frome2013devise,mao2015deep,wei2021universal} and \textit{multi-modal retrieval}~\cite{balaneshin2018deep,guo2018multi,laenen2018web}. 
In broad terms, multimodal retrieval refers to requests containing queries or results with modalities beyond text (e.g., images, videos, or audio).
Specially, cross-modal retrieval usually involves unimodal queries and results but with different modalities, e.g., text-to-image retrieval~\cite{frome2013devise,wang2019camp,socher2014grounded}, text-to-video retrieval~\cite{liu2018attentive,wang2022siamese}, image-to-caption retrieval~\cite{frome2013devise,mao2015deep,hodosh2013framing}, etc.
~\citet{balaneshin2018deep} jointly model the text-image embedding space by cross-modal alignment and unified representation learning.
Their approach can handle both cross-modal (text-to-image, T $\rightarrow$ I) and multimodal tasks (I $\rightarrow$ IT, T $\rightarrow$ IT).
Besides \textit{single query sessions}, researchers also aim at optimizing \textit{multi-query sessions}~\cite{liu2022pretraining,xie2019improving}.
For example, ~\citet{xie2019improving} enhance image search by leveraging session contexts such as historical queries and engaged images.
In addition, ~\citet{liu2022pretraining} utilize a heterogeneous graph neural network (HGN) to model intra-query, inter-query, and inter-modality information diffusion in multi-query product search.
Experimental results have shown the usefulness of session-level contexts in user intent modeling and multimodal information matching.

%A certain proportion of existing studies primarily focus on factoid retrieval tasks, characterized by clear intent and precise answers.
%These tasks typically have best matches for queries like ``a picture of a person riding a horse'' or an image of a red Adidas sneaker.

\subsection{Heterogeneous User Behavior Analysis}
Analyzing heterogeneous user behavior and further exerting specific patterns in corresponding scenarios is essential for user satisfaction modeling and system optimization.
To investigate user behavior in vertical search scenarios, researchers conduct in-depth log analysis for visual search~\cite{dagan2021image} or eye-tracking study for image search~\cite{xie2017investigating}, respectively.
Their findings help improve the corresponding ranking algorithms and system layout.
Besides, numerous studies have explored users' intent transition while interacting with a retrieval system~\cite{chen2021towards,wu2017returning,shi2024unisar}. 
For example, ~\citet{chen2021towards} investigate differences in user query reformulation behavior from delicate aspects such as the reformulation reason, interface, and the inspiration source via a field study.
While directly modeling user retention is challenging in practice, ~\citet{wu2017returning} discover that frequent revisits could indicate a stronger user stickiness.
Although these studies present valuable insights for optimizing the system, they mainly discuss a single component in the S\&R system.
As one service can significantly influence user behavior in another, e.g., a certain number of search queries rise from the suggestions in a recommended result, we collect APP-level user sessions for \textsf{Qilin}.
These sessions contain user feedback on search, recommendation, and DQA services, along with contextual information such as session IDs, search sources, and timestamps, to better analyze user tracks within and across services.

\subsection{Datasets for Search and Recommendation}
Datasets are the foundation of both improving and evaluating retrieval systems.
So far, most widely used datasets for search~\cite{nguyen2016ms,xie2023t2ranking}, recommendation~\cite{gao2022kuairec,marlin2009collaborative}, and S\&R~\cite{sun2023kuaisar,ai2017learning,liu2023jdsearch} usually only contain textual contents or value-based features for items, which is deficient for building better multimodal retrieval systems.
One exception is the e-commerce scenario~\cite{ren2024information,ai2017learning}, where the system needs to rank the products with both titles and images.
In this respect, ~\citet{miao2020multi} and ~\citet{gong2019automatic} present datasets containing short titles and images for Taobao products.
Besides the mentioned datasets, multimodal retrieval datasets such as UniIR~\cite{wei2025uniir} and Flickr30K~\cite{plummer2015flickr30k} have also been developed. 
These datasets mainly contain factoid queries such as ``find a picture of a person riding a horse'' which are characterized by clear intent and typically have ground truth matches.
However, in practical S\&R scenarios, user intents tend to be ambiguous.
Although these datasets are valuable for image-text modality alignment, they fall short when directly applied to the training of a general retrieval system.
Such a system requires the ability to capture complex user intents behind both factoid and non-factoid queries and retrieve pertinent results correspondingly.

In this regard, the \textsf{Qilin} dataset features a substantial number of non-factoid queries paired with multimodal results, which is quite challenging.
Moreover, collected user sessions contain abundant textual and image contents, facilitating the application of all aforementioned approach groups along with the investigation of complicated retrieval scenarios.
This complexity poses challenges for not only multimodal intent understanding but session-level cross-modal matching as well.

\section{The Qilin Dataset}
In this section, we delve into the details of the \textsf{Qilin} dataset.
We first elaborate on the data preprocessing and construction pipeline.
Then, a data schema of \textsf{Qilin} is presented to provide a clear glance at the information it includes.
Finally, we discuss the potential research tasks that \textsf{Qilin} may support in various scenarios such as search, recommendation, retrieval-augmented generation, etc.

\begin{figure}[t]
\centering
\includegraphics[width=\linewidth]{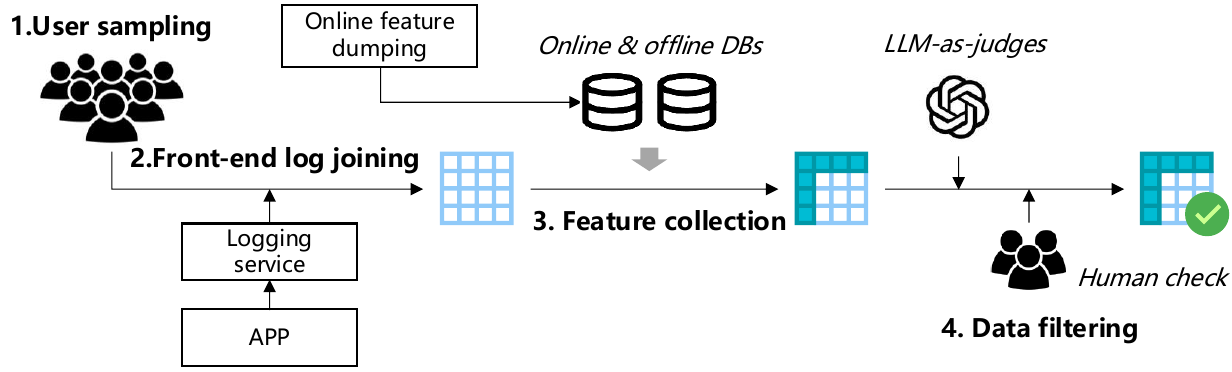}
\vspace{-0.2cm}
\caption{The data construction process of \textsf{Qilin}. The front-end log is joined with sampled user IDs to obtain the dataset backbone. Then we collect features for the request, user, and note from various databases. Finally, all content features undergo rigorous filtering by LLMs and human experts.}
\label{data_construct}
\end{figure}

% 可以给一个数据采样的流程图，不要算法图了看着难受
\subsection{Data Construction}\label{sec:data_construction}

As shown in Figure~\ref{data_construct}, the overall data construction pipeline of \textsf{Qilin} includes several steps: user sampling, front-end log joining, feature collection, and data filtering.\\
\textbf{User sampling.} 
To contain as much APP-level contextual information as possible, we randomly sample from the raw data w.r.t. the user IDs rather than the page views (PVs). 
All users are categorized into an engagement level according to their interaction frequency.
We first sample 15k non-spam users from the total pool of users who exhibited core interactive behaviors within the APP on November 27th, 2024.
These users show diverse behavioral patterns while interacting with the APP.
However, most of the above search requests did not trigger the DQA module.
To support intensive investigations concerning DQA, we select about 3,000 users who explicitly engaged with the module on the same date.
Finally, we obtain a merged list with 17,576 user IDs.
\\
\textbf{Front-end log joining.} 
Next, we join a specific user ID with the front-end log to obtain request-level information and user feedback.
Here we only use the log on the same date to control the final data size.
For each recommendation request, we collect all exposed results with their positions, timestamps, and five user feedback signals such as click, like, comment, etc.
As partial results are video notes, we also record viewing time as an implicit feedback label.
Besides the above information, a search request also contains a query and a source of the query keyword.
The query source records the interface type from which the current query originates.
For special requests with the exposure of a direct DQA answer, we reserve the answer content, the referred result note IDs, and four feedback labels (e.g., whether the answer is liked by the user). 

All the requests can be identified by a session ID, a request ID, and a user ID, where a session contains all requests from the APP being opened to closed through a user.
Based on these identifiers, interactive signals, interfaces, and timestamps, we can easily recover APP-level user tracks and further analyze their behavioral patterns across multiple services. 
Furthermore, contextual information such as exposure positions enables the investigation of unbiased learning to rank algorithms~\cite{ai2018unbiased,chen2023thuir} and click models~\cite{chuklin2022click,chen2020context,lin2024clickprompt}.\\
\textbf{Feature collection.} 
Requests and feedback labels composite the backbone of our dataset.
Next, we join the requests with note feature and user feature tables to support the training of sophisticated search or recommendation models.
For each note, we contain fields such as title, content, cover, and non-cover image IDs.
Note that there is a proportion of video notes (about 20\verb|~|40\%), we extract the key frames of a video as the image data to compress the final data size and to protect user privacy.
In addition, we also include other basic information such as note type, video duration, and multi-granularity taxonomies, as well as 30 statistically dense features like impression count, click count, etc.
As for all users, demographic features (e.g., gender, age) and 40 encrypted dense features are provided.
To further enhance user modeling in session search or context-aware S\&R ranking, we additionally collect 20 recently clicked note IDs before users initiate a specific request.
\\
\textbf{Data Filtering.} 
To efficiently leverage our data while maintaining its security, the filtering algorithm should be capable of excluding clearly unsafe content while preserving as much valid information as possible.
Therefore, we prompt LLMs with detailed instructions to classify document safety based on the title and the content.
The models we applied for text and image filtering are Qwen2.5-14B-Instruct~\cite{yang2024qwen25} and Qwen2-VL-7B-Instruct~\cite{yang2024qwen2technicalreport}, respectively.
The latter is a multimodal LLM that supports images as input and has achieved state-of-the-art performance on various visual understanding benchmarks.
As shown in Table~\ref{llm_prompts}, we point out four types of textual content that should be rigorously excluded.
The prompt used for image filtering is similar but with more rules.
Besides the four previously mentioned types, we also instruct the LLM to identify whether an image contains real human faces.
The LLM should not only provide a safety label but also a short textual description of the detected image.
Firstly, we filter out all the images with one or more portraits according to the output labels.
%As we have prompted the LLM not to be too strict, there are some false negatives.
To further protect user privacy, we use another text-only LLM with 14B parameters to classify the remaining images based on the descriptions.
Subsequently, we select a subset of 3,000 images and recruit three experts to double-check whether this subset includes any minor unsafe content.
Such a process is repeated several times until the manual verification is passed.
As a result, we obtain a corpus with 1,983,938 notes and 5,006,181 images.

\begin{table}[t]
\centering
\caption{Prompts used for filtering textual and image data.}
\vspace{-0.2cm}
\label{llm_prompts}
\begin{small}
\begin{tabularx}{\linewidth}{|X|}
\hline
\textbf{MODEL}: Qwen2.5-14B-Instruct (Qwen2-VL-7B-Instruct)\\
\hline
\textbf{PROMPT\_PREFIX}: 

Imagine you are a content safety reviewer for text (or images). Please examine the following text (or image) and assess whether it contains any illegal or sensitive information:

\textit{1. Pornographic/Obscene}: Includes explicit sexual descriptions, obscene language, etc.;

\textit{2. Violence}: Includes bloody, terrifying, extreme violence, etc.;

\textit{3. Political Figures}: Mentions or discusses information related to political figures;

\textit{4. Private Information of Ordinary Individuals}: Involves privacy such as names, identifiers, contact information, addresses, etc.;

\textit{5. Portraits}: Includes any real human faces. Note that the face of a virtual character (e.g., animation, comic, or game roles) is not a portrait;

\textbf{Additional Notes}:

$\star$ Advertising content does not count as illegal or sensitive information.

$\star$ Mentioning public figures (such as actors, singers, entrepreneurs, etc.) does not count as illegal or sensitive information unless they are political figures.

$\star$ Return True only if the text clearly contains any of the specified illegal or sensitive content; otherwise, return False.

$\star$ Any emoji is not pornographic content.

$\star$ Normal romantic-related content or sexual education content is not considered pornographic; but if the content is overly explicit or detailed, then it counts.

$\star$ Expressions of negative emotions, when not coupled with violent content, should not be classified as acts of violence.

$\star$ Do not be too strict, as minor issues being marked as illegal or sensitive can result in false positives. 
If you are uncertain, it is acceptable to refrain from labeling the content as illegal or sensitive. 
Only when you are very sure, mark it as a positive instance and return True.

Text (Image) to be checked: [\\% ！！！
\hline
\textbf{PROMPT\_SUFFIX}:

] Please only return True or False without any explanations.
\\
\hline
\end{tabularx}
\end{small}
\end{table}

\subsection{Statistics \& Data Schema}
Generally, \textsf{Qilin} comprises APP-level sessions from 15,482 users.
Comparison of basic statistics between \textsf{Qilin} and existing S\&R datasets (Amazon~\cite{he2016ups,mcauley2015image}, JD Search~\cite{liu2023jdsearch}, KuaiSAR~\cite{sun2023kuaisar}) is given in Table~\ref{comparison_statistics}.
Among these datasets, only Amazon can be marginally adopted for studying multimodal S\&R systems.
However, it only offers pseudo queries derived from the product metadata, which might compromise the reliability of experimental findings due to the absence of real user search behaviors.
Moreover, product titles and images in the Amazon dataset need to be crawled additionally, increasing the experiment cost.
On the other side, JD Search and KuaiSAR merely provide anonymized item contents (i.e., encrypted word ID sequence) which may cause difficulties in interpreting model effectiveness.
Fortunately, \textit{Xiaohongshu} maintains an open community with abundant user-generated content (UGC), aiming at facilitating human connection and mutual assistance.
To ensure the comprehensiveness of our dataset, we release the original note content (title + main body + images) after a thorough filtering process.

\begin{table}[t]
\centering
\caption{Comparison between \textsf{Qilin} and existing S\&R datasets. Note that queries in Amazon are pseudo ones. JD Search and KuaiSAR only provide anonymized item content. Besides S\&R, \textsf{Qilin} also include user actions on DQA.}
\vspace{-0.2cm}
\label{comparison_statistics}
\begin{threeparttable}
\begin{small}
\begin{tabular}{l|cccc}
\Xhline{0.7pt}
\textbf{Property} & \textbf{Amazon} & \textbf{JD Search} & \textbf{KuaiSAR} & \textbf{Qilin}\\
\hline
\# Users & 192,403 & 173,831 & 25,877 & 15,482\\
\# Items & 63,001 & 12,872,736 & 6,890,707 & 1,983,938 \\ 
\# Queries & 3,221 & 171,728 & 453,667 & 57,188$^{1}$\\
\# Actions & 1,689,188 & 26,667,260 & 19,664,885 & 2,498,594\\
\# Images$^{2}$ & ? & -- & -- & 5,006,181 \\
%Item image & $\surd$ & $\times$ & $\times$ & $\surd$\\
DQA info & $\times$ & $\times$ & $\times$ & $\surd$\\
Item text & title+review & anon'd & anon'd & title+body\\ % 明天问下搞推荐的同学
\Xhline{0.7pt}
\end{tabular}
\end{small}
\begin{tablenotes}
\footnotesize
	\item[1] \textsf{Qilin} provides a subset of 2,932 queries triggering the DQA module.
	\item[2] Images of Amazon need to be crawled on demand, thus no statistics here.
\end{tablenotes}
\end{threeparttable}
\end{table}

\begin{table}[t]
\centering
\caption{Data schema of \textsf{Qilin}.}
\vspace{-0.2cm}
\label{data_schema}
\begin{threeparttable}
\begin{small}
\begin{tabularx}{\linewidth}{llX}
\Xhline{0.7pt}
\textbf{Table} & \textbf{Key} & \textbf{Fields}\\
\hline
Search & search id & query, session id, user id, 20 recently clicked note ids, query source (1-8), search result details (list, each element is encapsulated as \{\textit{note id}, \textit{position}, \textit{timestamp}, \textit{six engage labels}\}, sorted by ascending timestamp), DQA details (if triggered);\\
\hline
Rec & request id & session id, user id, 20 recently clicked note ids, recommendation result details (a list of encapsulated elements, ditto);\\
\hline
DQA & search id & all search fields, answer content, referred note ids, four user engage labels;\\
\hline
User & user id & gender, platform, age, fan number, follow number, 40 encrypted dense features;\\
\hline
Note & note id & note type, note title, note content, image id list$^1$, video duration, video height, video width, image num, content length, commercial flag, 1/2/3-level taxonomy ids, 30 statistical features (e.g., the number of monthly impressions or clicks);\\
\Xhline{0.7pt}
\end{tabularx}
\end{small}
\begin{tablenotes}
\footnotesize
%	\item[1] Page time label is only valid for video notes.
	\item[1] For video notes, we select images by equidistant sampling from key frames.
\end{tablenotes}
\end{threeparttable}
\end{table}

%\textsf{Qilin} provides a subset of 2,888 queries triggering the DQA module.

% session num, request num, user num, note num, avg image num
\textsf{Qilin} is organized into tabular format as shown in Table~\ref{data_schema}.
The first three rows are the backbone tables, which contain the basic information for each search or recommendation request.
Besides the user ID, we also collect 20 recently clicked note IDs to support short-term user modeling.
Unlike existing missing-at-random datasets~\cite{gao2022kuairec}, we remain all exposed results at various positions for both search and recommendation.
When we know whether an item has been examined, then there is no need to preserve a high density of user-item interaction.
Binary feedback labels include click and four engaging actions: like (clicking the heart icon in Figure~\ref{app_tracks}), collect (clicking the star icon in Figure~\ref{app_tracks}), comment, and share.
Since these engaging actions are sparse, we hereby consider the viewing time in seconds as a label for model training.

For a search request triggering the DQA module, we record the content of the answer, the referred notes, and corresponding user feedback.
As direct answers are presented at the top of the result page, we assume all of them have been browsed by users.
To further distinguish subtle differences in user satisfaction towards the answer, we collect four user actions on the DQA module: 
1) whether the answer is liked by the user, 
2) whether the user clicks the reference superscript, 
3) whether the user clicks the answer body, and 
4) whether the user clicks the aggregated experience tab (to unfold the relevant passages contributing to the final answer).
Given referred notes and user-favored answers, we can easily train an LLM within an RAG pipeline and evaluate its performance.
Furthermore, researchers can also investigate the differences of user behaviors in search scenarios with and without the DQA module.

To facilitate the training of various approaches, we collect rich content-based and ID-based side features for users and notes.
Retrieval approaches in industry can be broadly divided into two groups.
The first group of approaches usually has a pyramid architecture, with enormous sparse ID embedding parameters but light weights for feature interaction, e.g., CTR models such as DCN-V2~\cite{wang2021dcn}.
These models are usually fast in convergence and have high inference speed.
However, they are poor in generalization due to the shallow semantic representation.
On the contrary, content-based approaches only involve word embeddings and exploit deep networks for representation learning and feature interaction, e.g., pre-trained models~\cite{fan2022pre}.
They are more robust to unseen data while requiring huge online computation resources for efficient deployment.
Considering the above issues, we emphasize the significance of combining content with sparse features to construct better retrieval systems.
For all notes, we attach an image list in chronological order, i.e., the first image in the list is the cover, which has a great influence on user clicking behaviors.
Since video contents are informationally redundant, we equidistantly sample images from key frames besides the cover.
All images are compressed in WebP format, which typically achieves an average of 30\% more compression than JPEG without loss of image quality.

\begin{table}[t]
\centering
\caption{Potential tasks that \textsf{Qilin} supports.}
\vspace{-0.2cm}
\label{potential_task}
\begin{small}
\begin{tabularx}{\linewidth}{l|X}
\Xhline{0.9pt}
\textbf{Scenario} & \textbf{Tasks}\\
\hline
Search & CTR prediction, click simulation, unbiased learning to rank, multimodal search, context-aware ranking, session search, (post) pre-training for web search, enhancing search by recommendation, pre-training for RAG, evaluation for RAG, multimodal RAG, query performance prediction, etc;\\
\hline
Rec & content-based recommendation, multimodal recommendation, session-based recommendation, unbiased recommendation, query (intent) recommendation, enhancing recommendation by search, etc;\\
\hline
General & multi-task learning, multi-scenario learning, heterogeneous user behavior analysis,  multimodal alignment, multimodal LLM-as-judges, scaling laws fitting, etc;\\
\Xhline{0.9pt}
\end{tabularx}
\end{small}
\end{table}

\subsection{Potential Tasks}
 Based on the features and information available, we present in Table~\ref{potential_task} the range of tasks that \textsf{Qilin} potentially supports.
Firstly, \textsf{Qilin} enables content-based retrieval and reranking for both search and recommendation.
By incorporating images into user intent or note content encoding, there is further space to explore multimodal search and recommender systems.
In heterogeneous result pages, there tend to be more user behavioral biases compared to traditional text-only pages.
As a result, unbiased learning to rank~\cite{ai2018unbiased,chen2023thuir} becomes crucial for improving system fairness and sustainability.
Next, by leveraging multiple user feedback labels across various scenarios,  it could be feasible to efficiently train multi-task or multi-scenario learning frameworks~\cite{shi2024unisar,ahmad2019context,chen2021hybrid}.
Besides joint learning, information from one scenario can be selectively applied to enhance another. 
For example, user actions in recommendation could be integrated into search ranking algorithms to better model user preferences~\cite{si2023search}.
As our dataset involves user engagements with different modules and services, there may be interesting heterogeneous user behavioral patterns that warrant further exploration.
Some other tasks include post pre-training for retrieval-augmented generation (RAG)~\cite{gao2023retrieval}, query performance prediction (QPP)~\cite{arabzadeh2021bert}, query (intent) recommendation~\cite{ozertem2012learning}, multimodal alignment, scaling laws fitting~\cite{fang2024scaling}, and using multimodal large language models as judges~\cite{li2024llms}.
Concretely, the LLM-based data filtering process in Section~$\S$\ref{sec:data_construction} can be regarded as an example of LLM-as-judges.

Last but not least, \textsf{Qilin} can be taken as a semi-finished benchmark when augmented with precise human annotations to support more tasks like user satisfaction modeling, entity-enhanced search, multimodal RAG evaluation, etc.

\section{Primary Data Analysis}
In this section, we present primary data analysis for \textsf{Qilin}, mainly including four parts: 1) demographics, 2) engagement \& result distribution, 3) transitions across services, and 4) query analysis.

% demographic
\subsection{Demographics}
According to IP locations, 15,482 users come from more than 87 countries or regions, mainly including China (84.32\%), Andorra (1.45\%), Australia (1.22\%), Iceland (1.19\%), Malaysia (1.14\%), Japan, South Korea, United States, United Kingdom, Canada, Ireland, France, Singapore, Austria, and Germany.
76.53\% of them are female, while others are male, with a ratio slightly higher than the average level of the APP.
This may be because we retain a certain proportion of search requests that trigger the DQA module, and women users tend to engage more within these requests.
Most users are young and middle-aged individuals, in the age range of 16 to 40 years old (74.5\%).
There are also small proportions of users below 16 (3.88\%) and over 40 (7.27\%).
As for the platform, iOS and Android accounts for 53.56\% and 38.79\%, respectively.
The rest of the users issue requests from desktop websites or other mobile operating systems such as Harmony.

% examination & user feedback
\subsection{Engagement \& result distribution}
Based on the \textsf{Qilin} dataset, in this section we aim to analyze 1) user engagements across scenarios, 2) user behavioral biases, and 3) result type distributions.

\begin{table}[t]
\centering
\caption{User engagements across scenarios, where S-DQA and S+DQA denote search requests without or with triggering the DQA module. Note that engagement rates such as like rate and collect rate are calculated based on the conditional probability $P(engage=1|click=1)$.}
\vspace{-0.2cm}
\label{user_engagement}
\begin{small}
\begin{tabular}{l|cc|cc}
\Xhline{0.9pt}
\textbf{} & \textbf{S} & \textbf{R} & \textbf{S-DQA} & \textbf{S+DQA}\\
\hline
\textit{Avg. browsing depth} & \cellcolor{mygray}{22.75} & 18.80 & \cellcolor{mygray}{23.41} & 10.61\\
\textit{Avg. first click rank} & 3.01 & \cellcolor{mygray}{4.97} & \cellcolor{mygray}{3.03} & 2.50\\
\textit{Avg. click num} & \cellcolor{mygray}{3.88} & 3.67 & \cellcolor{mygray}{3.99} & 2.50\\
\hline
\textit{Click-through rate} & 21.01\% & \cellcolor{mygray}{24.13\%} & \cellcolor{mygray}{21.02\%} & 20.73\%\\
\textit{Like rate} & 4.11\% & \cellcolor{mygray}{7.07\%} & \cellcolor{mygray}{4.19\%} & 1.29\%\\
\textit{Collect rate} & \cellcolor{mygray}{1.87\%} & 1.47\% & \cellcolor{mygray}{1.88\%} & 1.26\%\\
\textit{Share rate} & 0.57\% & \cellcolor{mygray}{0.63\%} & \cellcolor{mygray}{0.57\%} & 0.52\%\\
\textit{Comment rate} & 0.32\% & \cellcolor{mygray}{0.99\%} & \cellcolor{mygray}{0.32\%} & 0.21\%\\
\hline
\textit{\# Samples} & 57,188 & 94,552 & 54,256 & 2,932\\
\Xhline{0.9pt}
\end{tabular}
\end{small}
\end{table}

To explore user engagements, we consider two scenario pairs for comparison: S (Search) vs. R (Recommendation) and S+DQA (Search with DQA) vs. S-DQA (Search without DQA).
As shown in Table~\ref{user_engagement}, we have several findings.
Firstly, compared to search, users click fewer results but have higher engagement rates except for the collecting action in the recommendation.
With clearer intents, users may browse more results to find relevant or useful notes in the search service, resulting in a high browsing depth and click number.
Once users find a helpful note, they may save it to the private collection for future demands.
Overall, in the recommendation process, user intents are diverse and ambiguous.
They may just click on several funny notes to kill time and usually scroll from one video to another.
Therefore, recommendation users may browse fewer results on the original result page while browsing and liking more results in the embedded screaming.
By comparing the right two columns, we find that the DQA module can greatly impact user behaviors.
When provided with the direct answer, users only click several high-related notes at the top (i.e., with a lower first click position) and engage with significantly fewer notes.
There could be two reasons: 
1) Queries that trigger the DQA module tend to be factoid questions, where user information needs can be easily satisfied by a small number of notes.
2) By providing a direct answer with reference notes, the DQA module can save user effort in searching and summarizing useful information. 
Given this significant difference in user behavior, further exploration is required on how to rank the original results and design the result page layout to enhance users' search experience.

\begin{figure}[t]
\centering
\subfigure[Position bias.]{
\label{position_bias} %% label for first subfigure
\includegraphics[width=4.1cm]{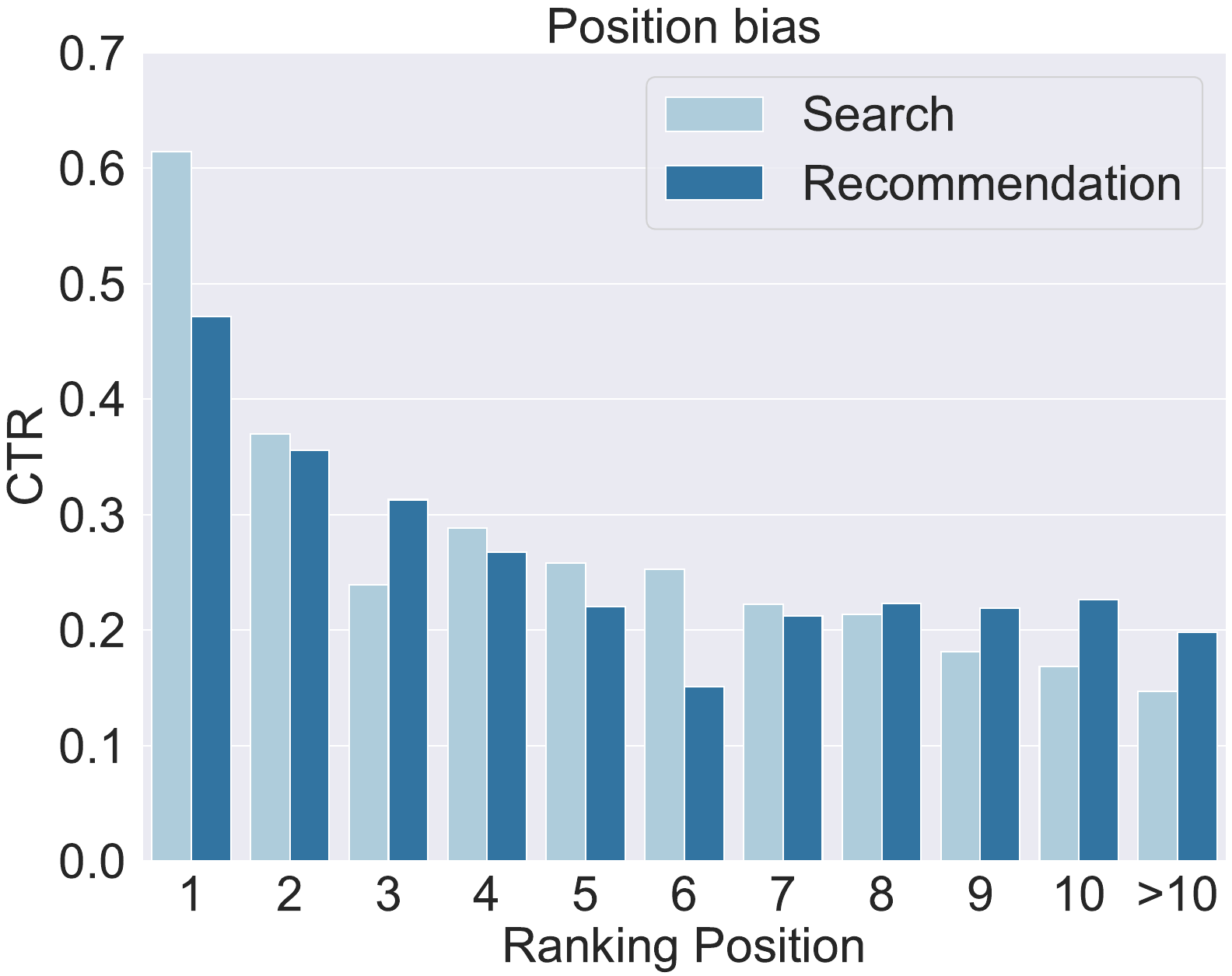}}
\subfigure[Session bias.]{
\label{session_bias} %% label for first subfigure
\includegraphics[width=4.1cm]{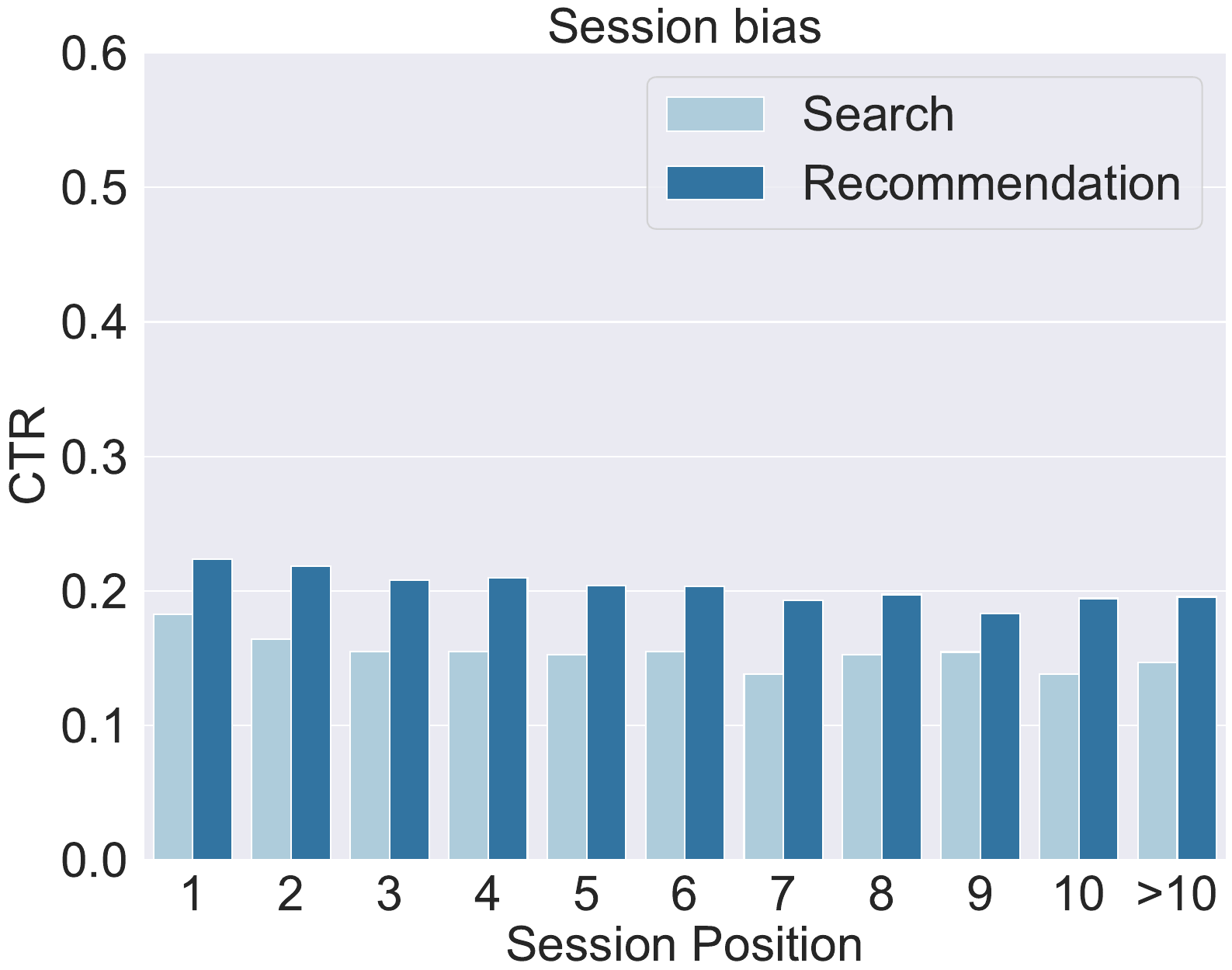}}
\subfigure[Search result distribution.]{
\label{search_result} %% label for first subfigure
\includegraphics[width=4.1cm]{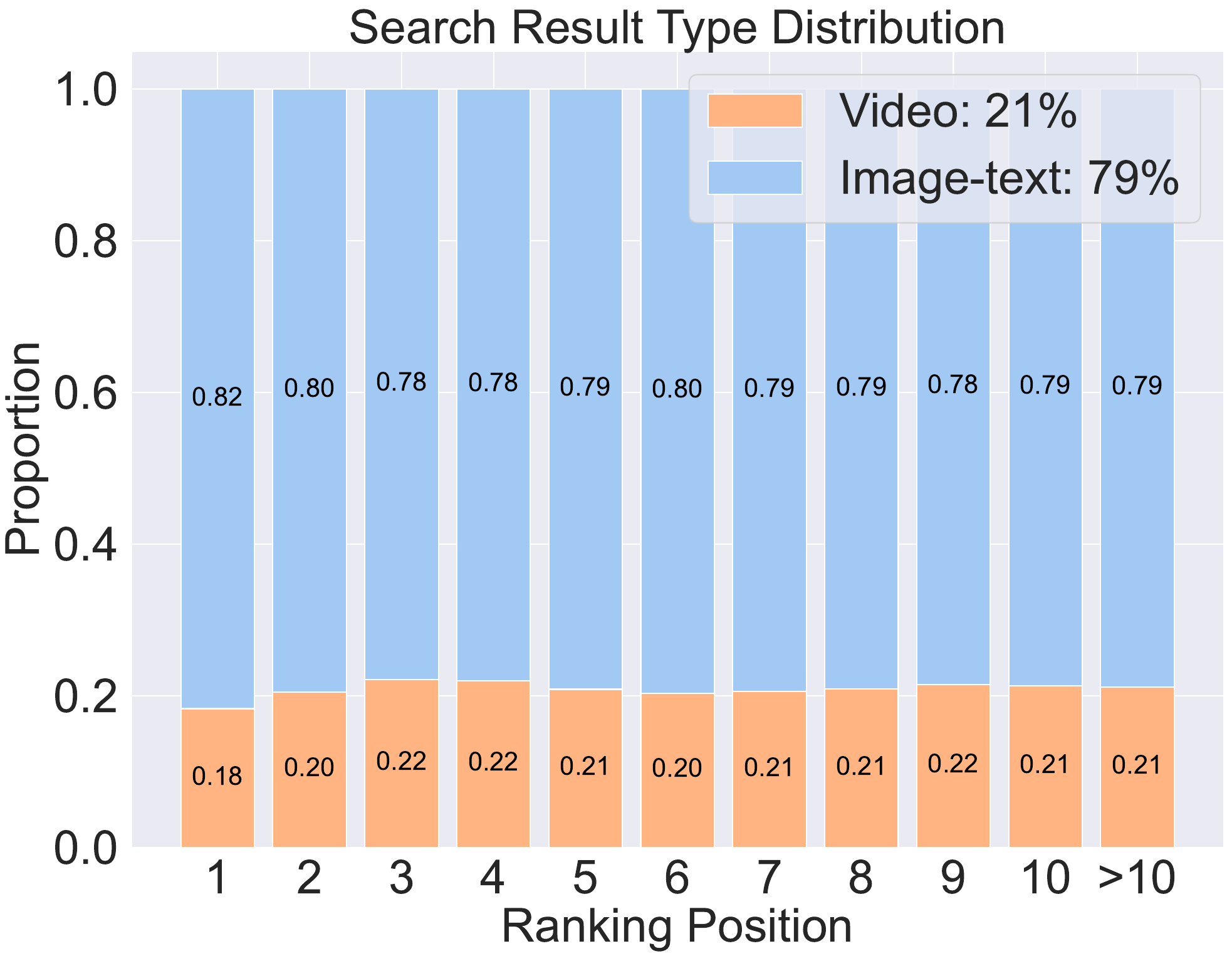}}
\subfigure[Recommendation result distribution.]{
\label{rec_result} %% label for first subfigure
\includegraphics[width=4.1cm]{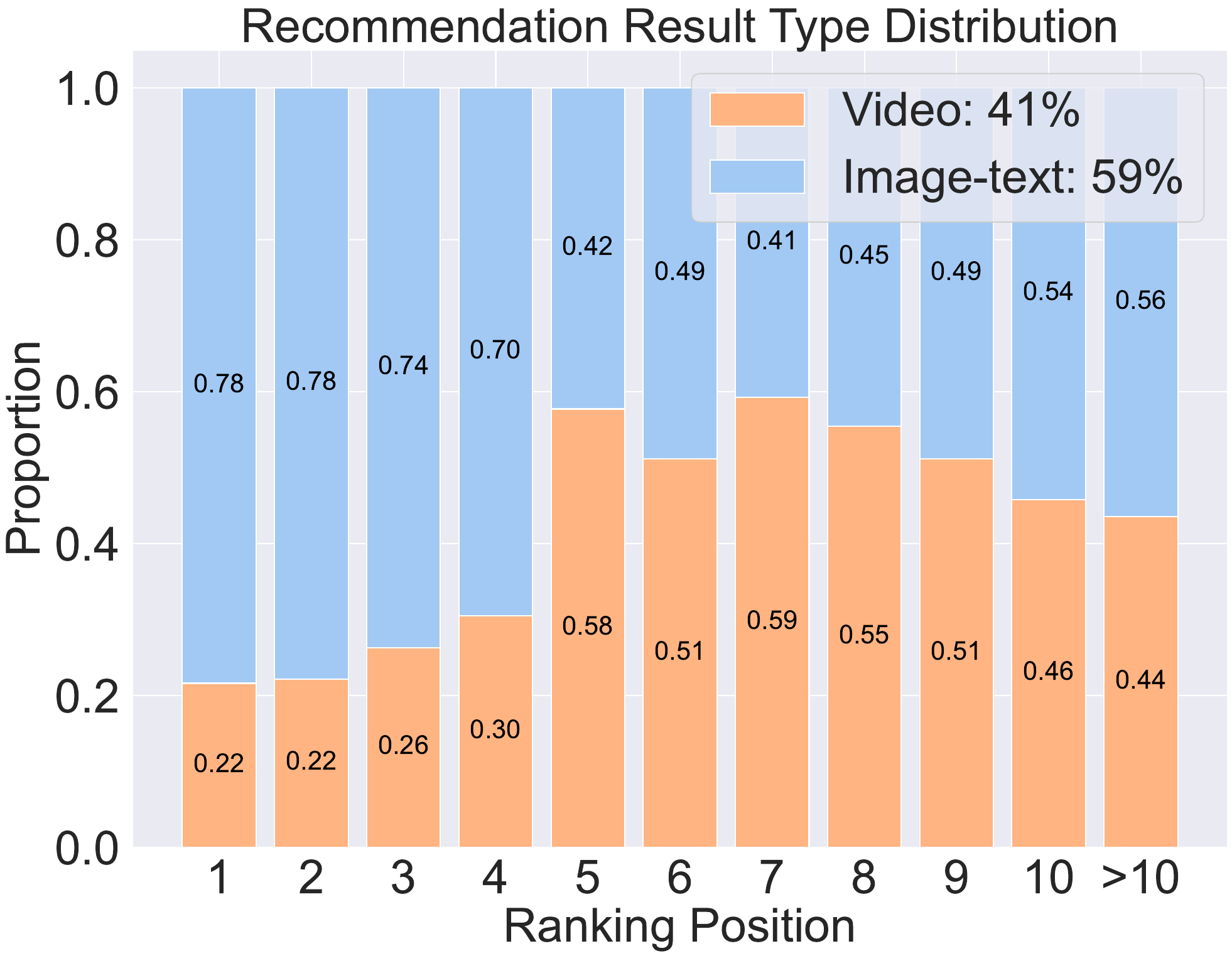}}
\subfigure[Search result CTR.]{
\label{search_result_ctr} %% label for first subfigure
\includegraphics[width=4cm]{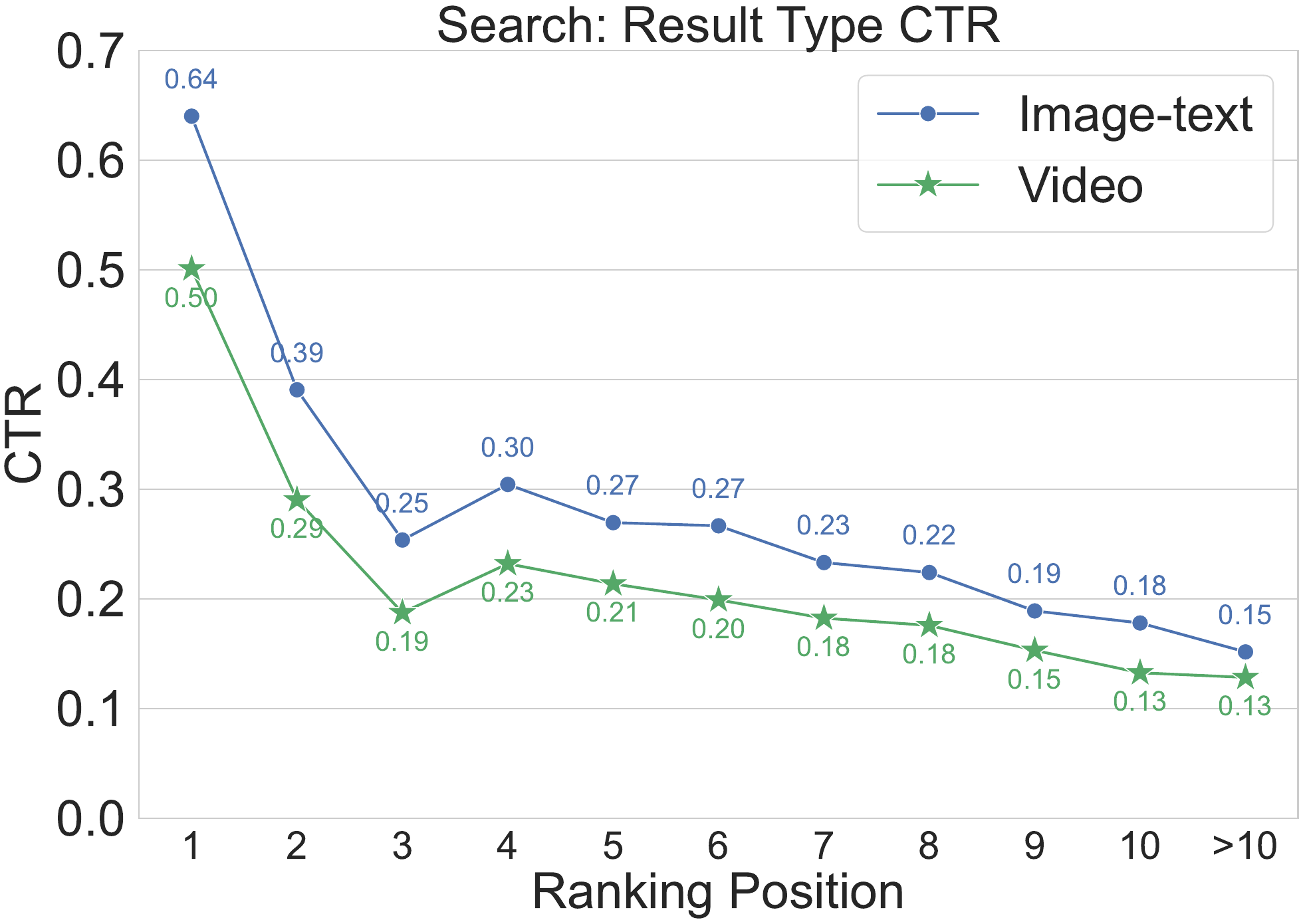}}
\subfigure[Recommendation result CTR.]{
\label{rec_result_ctr} %% label for first subfigure
\includegraphics[width=4cm]{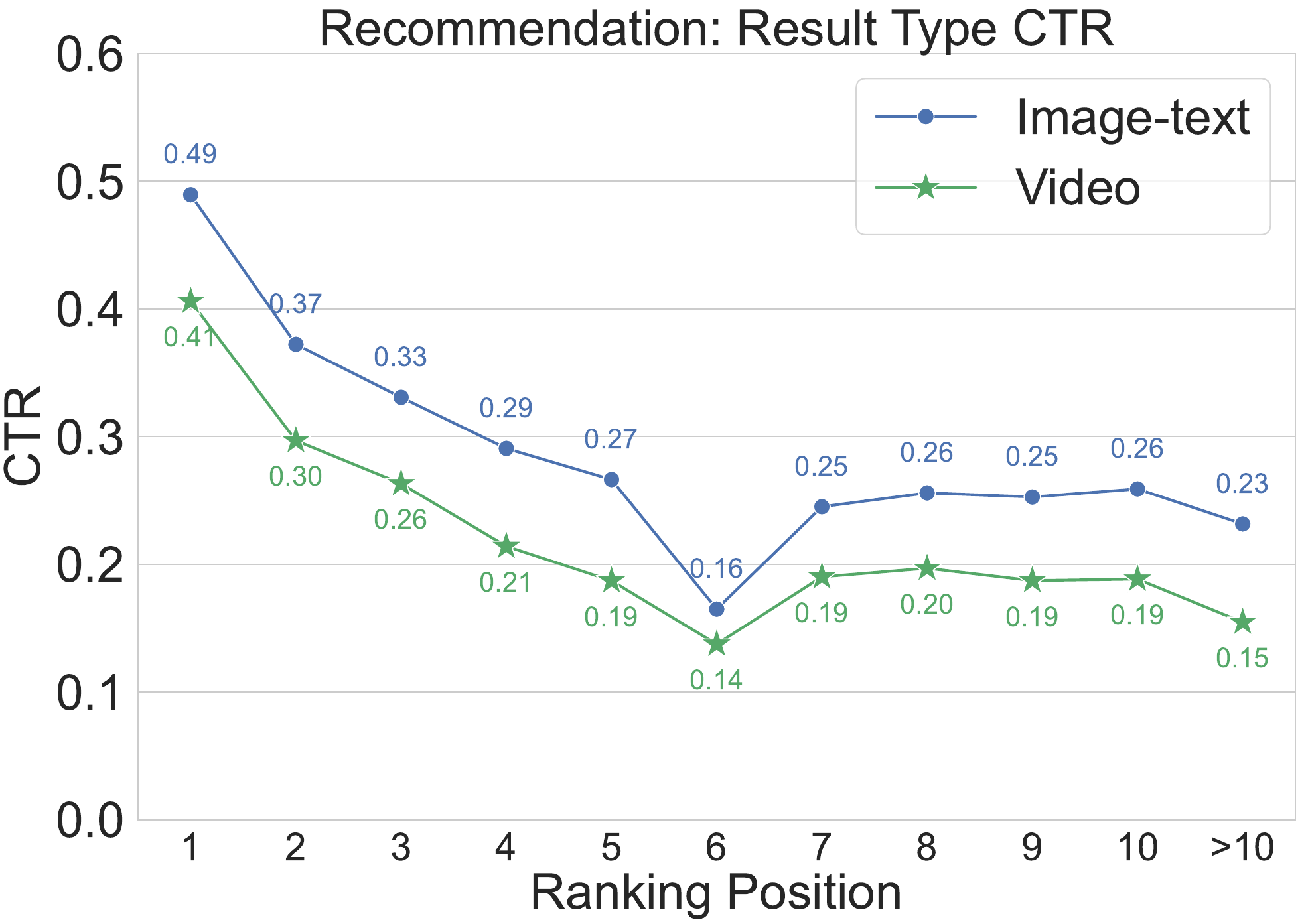}}
\vspace{-0.2cm}
\caption{Position bias, session bias, result type distribution, and CTR for two result types w.r.t. ranking positions in search and recommendation scenarios.}
\label{bias_result}
\end{figure}

Next, we plot CTR across ranking positions and session positions in Figure~\ref{position_bias} and Figure~\ref{session_bias}, respectively.
For both search and recommendation, there is a decay in CTR when the ranking position increases, which is often referred to as the position bias.
The decay is relatively slower in recommendation, suggesting that user browsing behavior may be more random compared to search, where users are more likely to click on top results.
It is quite intriguing to observe a dip in the distribution at the third and the sixth positions for search and recommendation.
We guess these positions have higher probabilities of exposing commercial notes~\footnote{A commercial note is an image-text note or a video note with a product or advertisement, only accounting for a tiny proportion of the data.}, thus users are not very inclined to click on them.
For the session dimension, we find that there also exists a decaying click rate when users issue more requests within an APP-level session.
However, this decay is more gradual at the session scale.
As search users are in a trial-and-error process, the corresponding CTR curve decays faster at the end of a session compared to the recommendation.
Factors such as user fatigue or satisfaction may contribute to this session bias, i.e., users' accumulated cost or gain have reached the upper limit, so they are more reluctant to click on a result in the later stage of sessions.

For result types, we only consider image-text and video notes.
We first calculate the distribution proportion of each note type across ranking positions.
As revealed in Figure~\ref{search_result} and Figure~\ref{rec_result}, the proportion of video notes in both S\&R scenarios initially rises and then levels off.
By comparing the two figures, it is obvious video notes are exposed more in the recommendation.
This phenomenon may be caused by prolonged user-system in-loop interaction.
Firstly, users are more inclined to view videos to kill time when using recommendation services.
After some time, the recommender system may have learned this pattern and automatically ranks more videos at the top positions.
Users continue to click on these video notes, resulting in a huge difference in the exposure structure between the search and recommendation scenarios.
Generally, users have a higher CTR on image-text notes in both scenarios (see Figure~\ref{search_result_ctr} and ~\ref{rec_result_ctr}).
We guess there are two main causes: the difference in browsing effort and the limitation of the user environment.
Video notes typically require more time to complete viewing, contain less textual information per screen, and may produce noise in the surrounding environment.
These limitations make users more inclined to click on image-text notes in most circumstances.
However, video notes can offer greater advantages in certain situations, such as for entertainment or when there is a need for detailed moment-level information.
To this end, more investigation should be conducted on selecting the appropriate note type for specific user intents or request conditions to improve user satisfaction in the future.
\begin{table}[t]
\centering
\caption{User transition analysis. S and R stand for search and recommendation, respectively.}
\vspace{-0.2cm}
\label{user_transition}
\begin{small}
\begin{tabular}{l|cccc}
\Xhline{0.9pt}
\textbf{} & \textbf{S$\rightarrow$S} & \textbf{R$\rightarrow$S} & \textbf{R$\rightarrow$R} & \textbf{S$\rightarrow$R}\\
\hline
\textit{Proportion} & 34.88\% & 2.95\% & 59.32\% & 2.84\% \\
\textit{Avg. click num} & 4.0014 & 3.9782 & 2.9745 & 4.4466\\
\textit{Click-through rate} & 20.18\% & 22.98\% & 23.20\% & 25.01\%\\
\hline
\# Samples & 29,295 & 2,476 & 49,815 & 2,387\\
\Xhline{0.9pt}
\end{tabular}
\end{small}
\end{table}

\begin{table}[t]
\centering
\caption{Query length distribution and corresponding user engagement. Note that a deeper color denotes a higher value of CTR or average click number.}
\vspace{-0.2cm}
\label{query_length}
\begin{small}
\begin{tabular}{l|ccc}
\Xhline{0.9pt}
\textbf{} & \textbf{Proportion} & \textbf{CTR} & \textbf{Avg. click num}\\
\hline
\textit{Length<=3} & 0.32\% & \cellcolor{b4}{0.1967} & \cellcolor{b4}{2.9892}\\
\textit{Length=4} & 13.64\% & \cellcolor{b4}{0.1962} & \cellcolor{b2}{4.0276}\\
\textit{Length=5} & 13.94\% & \cellcolor{b4}{0.2047} & \cellcolor{b3}{3.9303}\\
\textit{Length=6} & 17.27\% & \cellcolor{b4}{0.2051} & \cellcolor{b1}{4.0338}\\
\textit{Length=7} & 14.07\% & \cellcolor{b3}{0.2111} & \cellcolor{b3}{3.9446}\\
\textit{Length=8} & 12.42\% & \cellcolor{b3}{0.2150} & \cellcolor{b3}{3.8180}\\
\textit{Length=9} & 9.54\% & \cellcolor{b2}{0.2199} & \cellcolor{b4}{3.6971}\\
\textit{Length=10} & 6.89\% & \cellcolor{b1}{0.2272} & \cellcolor{b4}{3.7497}\\
\textit{Length>10} & 11.91\% & \cellcolor{b2}{0.2216} & \cellcolor{b4}{3.7010}\\
\Xhline{0.9pt}
\end{tabular}
\end{small}
\end{table}

\begin{CJK}{UTF8}{gbsn}
\begin{table*}[t]
\centering
\caption{Examples of different query reformulation types. The total number of reformulating actions is 29,875.}
\vspace{-0.2cm}
\label{query_reformulation}
\begin{small}
\begin{tabular}{l|c|l}
\Xhline{0.9pt}
\textbf{Type} & \textbf{Prop.} & \textbf{Examples}\\
\hline
\textit{Add} & 13.57\% & \tabincell{l}{广州批发$\rightarrow$广州批发市场 (Guangzhou wholesale$\rightarrow$Guangzhou wholesale market)\\ 
%2) 宇宙科技画$\rightarrow$宇宙未来科技主题画 (cosmic technology art$\rightarrow$cosmic future technology-themed art); \\
五大文明$\rightarrow$五大文明发源地 (the five great civilizations$\rightarrow$the birthplaces of the five great civilizations)}\\
\hline
\textit{Delete} & 2.53\% & \tabincell{l}{ip设计插件$\rightarrow$ip设计 (intellectual property designing plugin$\rightarrow$intellectual property design)\\农业经营模式分析$\rightarrow$农业经营模式 (analysis of agricultural business modes$\rightarrow$agricultural business modes)} \\
\hline
\textit{Change} & 47.16\% & \tabincell{l}{文案破碎感$\rightarrow$文案自由松弛感 (writing sense with vulnerability$\rightarrow$writing sense with freedom and relaxation)\\将数据分三分位$\rightarrow$四分位间距 (dividing data into tertiles$\rightarrow$Interquartile Range, IQR)} \\
\hline
\textit{Repeat} & 7.57\% & \tabincell{l}{usmtekun美食$\rightarrow$usmtekun美食 (delicacy of usm tekun cafe, which refers to a cafe in Malaysia)\\脸部画法$\rightarrow$脸部画法 (techniques of drawing a face)} \\
\hline
\textit{Others} & 29.17\% & \tabincell{l}{相机镜头进灰$\rightarrow$适马56 (the camera len has dust in it$\rightarrow$SIGMA 56mm camera lens)\\挡脸怎么弄好看$\rightarrow$文案配照片 (cover the face in a visually appealing way$\rightarrow$match captions with photos)}\\
\Xhline{0.9pt}
\end{tabular}
\end{small}
\end{table*}
\end{CJK}

% \begin{table*}[t]
% \centering
% \caption{Examples of different query reformulation types. The total number of reformulating actions is 29,875.}
% \vspace{-0.2cm}
% \label{query_reformulation}
% \begin{small}
% \begin{tabular}{l|c|p{9cm}}  % 调整最后一列宽度
% \Xhline{0.9pt}
% \textbf{Type} & \textbf{Prop.} & \textbf{Examples}\\
% \hline
% \textit{Add} & 13.57\% & \makecell[l]{广州批发$\rightarrow$广州批发市场 \\ 
% 五大文明$\rightarrow$五大文明发源地} \\
% \hline
% \textit{Delete} & 2.53\% & \makecell[l]{ip设计插件$\rightarrow$ip设计 \\ 
% 农业经营模式分析$\rightarrow$农业经营模式} \\
% \hline
% \textit{Change} & 47.16\% & \makecell[l]{文案破碎感$\rightarrow$文案自由松弛感 \\ 
% 将数据分三分位$\rightarrow$四分位间距} \\
% \hline
% \textit{Repeat} & 7.57\% & \makecell[l]{usmtekun美食$\rightarrow$usmtekun美食 \\ 
% 脸部画法$\rightarrow$脸部画法} \\
% \hline
% \textit{Others} & 29.17\% & \makecell[l]{相机镜头进灰$\rightarrow$适马56 \\ 
% 挡脸怎么弄好看$\rightarrow$文案配照片} \\
% \Xhline{0.9pt}
% \end{tabular}
% \end{small}
% \end{table*}

% transition proportion: s->s, s->r, r->s, r->r;
\subsection{Transitions across services} \label{sec:transition}
This section presents user transitions across S\&R services.
As APP-level sessions may contain noises in user intent transformation, we redefine a session for more accurate transition analysis.
If the beginning timestamps of two requests (search or recommendation) from one user are within the 30-minute gap, then they are considered to belong to one session.
As shown in Table~\ref{user_transition}, users mainly transfer within search or recommendation services.
There is a certain proportion of transition from R$\rightarrow$S (2.95\%) and S$\rightarrow$R (2.84\%).
When transferred from recommendation, users tend to click search results with a higher click-through rate compared to the general search condition.
Similarly, users also have higher engagements with exposed results when transferring from search to recommendation, indicating the potential of considering these transition behaviors for better user preference modeling~\cite{shi2024unisar}.

% query analysis & reformulation analysis &  search source analysis
\subsection{Query analysis}
Following, we analyze search queries in the \textsf{Qilin} dataset, including character-based query length, query reformulation types, and search sources.
Firstly, we calculate the proportion of queries with different lengths and further compare user engagement under these query groups.
From Table~\ref{query_length}, we can observe that most queries contain 4-8 characters, under which users tend to click on more results (i.e., the middle-attention of the rightmost column).
These queries clearly express user intentions and are not too difficult, allowing the search system to return more relevant results.
Intriguingly, the click-through rate keeps rising as the query length increases.
Apart from the fact that longer queries tend to have clearer intent and better results to be matched, users are generally more eager to obtain satisfactory answers when issuing longer queries, e.g., task-oriented ones.
Therefore, they are more likely to click a search result under longer queries.

Secondly, we aim to explore users' query reformulation actions.
Following ~\cite{huang2009analyzing, chen2021towards}, we define five types of reformulating behaviors based on syntax relationships between two consecutive queries $q_{t-1}$ and $q_{t}$. 
Given $+q_t=\{w|w \in W(q_t), w \notin W(q_{q-1})\}$, $-q_t=\{w|w \notin W(q_t), w \in W(q_{t-1})\}$, and $\cap q_t=\{w|w \in W(q_t), w \in W(q_{t-1})\}$ where $w$ and $W(\cdot)$ denote a word~\footnote{As \textsf{Qilin} is Chinese-centric, we count a character as a word.} and the word set, we normalize the expression of Add, Delete, Change, Repeat, and Other reformulations as follows: 
\begin{flalign}
	& Add: +\Delta q_t \neq \emptyset, -\Delta q_t = \emptyset;\\
	& Delete: +\Delta q_t = \emptyset, -\Delta q_t \neq \emptyset;\\
	& Change: +\Delta q_t \neq \emptyset, -\Delta q_t \neq \emptyset, \cap q_t \neq \emptyset ;\\
	& Repeat: +\Delta q_t = \emptyset, -\Delta q_t = \emptyset, \cap q_t \neq \emptyset ;\\
	& Others: +\Delta q_t \neq \emptyset, -\Delta q_t \neq \emptyset, \cap q_t = \emptyset ;
\end{flalign}

Based on the above definitions and the S$\rightarrow$S sessions obtained from Section \S\ref{sec:transition}, we present the proportions of various syntactic query reformulation types and their corresponding examples in Table~\ref{query_reformulation}.
Type ``Change'', ``Add'' and ``Others'' account for about 90\% of all samples, which is consistent with previous findings in general search~\cite{chen2021towards}.
The balanced distribution of all types implies that users' search propensity in \textit{Xiaohongshu} is quite complex.
When user search intent is specialized or generalized, they may add or delete keywords based on the last query.
For example, having issued a query ``the five great civilizations'', the user may be curious about the origin of these civilizations and subsequently submit a second query.
When users are immersed in a spontaneous information-seeking flow, their trust and stickiness toward the system can be substantial. 
Therefore, it will be crucial to improve context-aware user search experiences for the whole S\&R system.

Besides query reformulations, query sources can also be used to model user state change.
According to the interface functionality, tens of entries can be broadly categorized into eight types, listed in Table~\ref{query_source}. 
We find that about half of the queries are from the auto-completion module, which is quite different from the general search engines such as Google and Baidu where users mostly issue a query via the input box by themselves~\cite{chen2021towards}.
About 6\% of queries come from history, indicating users may have similar search intents within a time interval.
Additionally, users will also click query suggestions presented in result pages or notes.
Modeling future intents and then recommending potential queries for users at an appropriate time can further improve user engagement.

\begin{table}[t]
\centering
\caption{Query source definition and analysis.}
\vspace{-0.2cm}
\label{query_source}
\begin{threeparttable}
\begin{small}
\begin{tabular}{l|l|l|c}
\Xhline{0.9pt}
 & \textbf{Name} & \textbf{Definition} & \textbf{Prop.} \\
\hline
1 & \textit{Active search} & using the search input box & 33.70\%\\
2 & \textit{Auto-completion} & using the auto-completion & 50.97\%\\
3 & \textit{Search history} & revisiting search history & 5.78\%\\
4 & \textit{Suggestion I} & clicking result page suggestions & 3.97\%\\
5 & \textit{Suggestion II}&  clicking suggestions in a result & 3.02\%\\
6 & \textit{Search filter} & using search filters$^{\dag}$ & 1.37\%\\
7 & \textit{Hotlist} & clicking on a hotlist & 0.63\%\\
8 & \textit{Others} & all other types & 0.55\%\\
\Xhline{0.9pt}
\end{tabular}
\end{small}
\begin{tablenotes}
\footnotesize
	\item[$\dag$] Including filters such as ``most recent notes'', ``video notes only'', and etc.
\end{tablenotes}
\end{threeparttable}
\end{table}

% 要不要做S&R的实验呢？
\section{Experiments}
In this section, we report the performance of baseline approaches on search, recommendation, and DQA tasks, respectively.
Due to the page limit, we do not consider jointly optimizing multiple tasks and leave it as future work. 

\subsection{Search and Recommendation}
We sort all exposed samples in chronological order and split them into training and testing sets with a ratio of 11:1 in hours, i.e., samples after 22:00 pm will be taken as the testing ones.
Baselines include BM25, BERT~\cite{devlin2018bert} bi-encoder, BERT cross-encoder, DCN-V2~\cite{wang2021dcn}, and Visual Language Model (VLM)~\cite{yang2024qwen2technicalreport}. 
In recommendation, we use the concatenated titles of recently clicked notes as the pseudo query to model user preference.
For BM25, BERT bi-encoder, and BERT cross-encoder, we consider query text, note title and note content as input.
%Additionally, VLM further integrates cover images of the current note and recently clicked into the feature interaction.
Additionally, VLM further integrates cover images into note encoding.
As for DCN-V2, besides the query and note (title + note) embeddings generated by pre-trained BERT bi-encoders (mean-pooled hidden states), we also utilize various dense features and contextual signals such as recently clicked notes provided in \textsf{Qilin}.
We scale up DCN-V2's trainable parameters to 0.13B, which is comparable with the BERT-base-chinese model (0.1B).
Except for the VLM, all these models are trained in an end-to-end manner.
We choose Qwen2-VL-7B-Instruct~\cite{yang2024qwen2technicalreport} with 4-bit quantization as the backbone of VLM and train it with LoRA~\cite{hu2021lora}, which is a parameter-efficient fine-tuning approach.
By setting lora\_rank to 16, only about 10M VLM parameters are tuned to ensure a fair comparison.
All the experimental details can be found in the released repository hereinabove.

\begin{table}[t]
\centering
\caption{Comparison of search and recommendation performances on various approaches.}
\vspace{-0.2cm}
\label{sar_exp}
\begin{small}
\begin{tabular}{l|c|c|c|c}
\Xhline{0.9pt}
\multirow{2}{*}{\textbf{Model}} & \multicolumn{4}{c}{\textbf{Search}} \\
\cline{2-5}
& MRR@10 & MRR@100 & MAP@10 & MAP@100\\
\hline
BM25 & 0.3388 & 0.3467 & 0.2399 & 0.3139 \\
BERT$_{bi}$ & 0.5320 & 0.5359 & 0.3855 & 0.4536 \\
BERT$_{cross}$ & 0.5336 & 0.5386 & 0.3848 & 0.4533 \\
%BERT4Rec & -- & -- & &\\
%DCN-V2 w/o Emb &  &  &  &  \\
DCN-V2 & \textbf{0.5600} & \textbf{0.5653} & \textbf{0.4014} & \textbf{0.4683} \\
VLM & 0.5523 & 0.5563 & 0.3937 & 0.4601\\
\hline
\multirow{2}{*}{\textbf{Model}} & \multicolumn{4}{c}{\textbf{Recommendation}} \\
\cline{2-5}
& MRR@10 & MRR@100 & MAP@10 & MAP@100 \\
\hline
BM25 & 0.5379 & 0.5418 & 0.3933 & 0.4634 \\
BERT$_{bi}$ & 0.6067 & 0.6087 & 0.4548 & 0.5183 \\
BERT$_{cross}$ & 0.6346 & 0.6362 & 0.4786 & 0.5394 \\
%BERT4Rec & -- & -- & &\\
%DCN-V2 w/o Emb &  &  &  &  \\
DCN-V2 & 0.6307 & 0.6321 & 0.4651 & 0.5278 \\
VLM & \textbf{0.6394} & \textbf{0.6409} & \textbf{0.4890} & \textbf{0.5477}\\ % !!
\Xhline{0.9pt}
\end{tabular}
\end{small}
\end{table}

For evaluation metrics, we choose Mean Reciprocal Rank (MRR) and Mean Average Precision (MAP), which can be formulated as:
\begin{flalign}
	MRR@K &= \frac{1}{N}\sum_{i}^{N}\frac{\mathbb{I}(r_i \leq K)}{r_i}\\
	MAP@K &= \frac{1}{N}\sum_{i}^{N}\frac{1}{R_i}\sum_{k}^{K} P_{i, k} \cdot \mathbb{I}(rel_{i, k}=1)
\end{flalign}
where $N$ denotes the total number of testing instances, $r_i$ represents the rank of the first positive result in the reranked list. 
$R_i$ and $P_{i, k}$ are the number of relevant results and the precision at $k$ metric for the $i$-th instance, respectively.
As for $rel_{i, k}$, we use clicks as binary relevance labels for simplicity.
%For DCN-V2, we consider two variants: incorporating pre-trained BERT bi-encoder embeddings for note titles and queries (DCN-V2) or not (DCN-V2 w/o Emb).

All results are shown in Table~\ref{sar_exp}.
For both search and recommendation, the BERT cross-encoder outperforms the bi-encoder, which is consistent with the findings of existing work.
Explicit query and document interaction help the model better learn the relevance matching.
By considering visual information, VLM further achieves better performance on two tasks, indicating the effectiveness of considering note images in both user modeling and note representing.
Compared with these pre-trained models, DCN-V2 shows competitive performance in both tasks.
It combines user history, sparse ID-based features, dense features, and pre-trained semantic embeddings altogether, thereby performing the best in search ranking.
However, the advantage of DCN-V2 is relatively smaller in the recommendation.
There may be two main reasons:
1) In our setting, the pseudo query we use in recommendation already summarizes user preference. Therefore, the margin between DCN-V2 and other approaches is narrowed.
2) Recommendation requires a higher model robustness to deal with the out-of-distribution problem. 
As DCN-V2 heavily depends on sparse features and lacks deep modeling of semantic signal matching, it may fall short in the recommendation.
Overall, the experimental results have shown the great potential of considering multimodal features and rich contextual signals to optimize the retrieval system.

\subsection{DQA}
For the DQA task, all user-engaged answers are considered the standard answer for evaluation.
We then adopt five popular LLMs (GPT3.5~\cite{achiam2023gpt}, GPT4o-mini~\cite{achiam2023gpt}, Qwen2.5-72B-Instruct~\cite{yang2024qwen25}, Llama3.3-70B-Instruct~\cite{dubey2024llama}, GLM4~\cite{glm2024chatglm}) to generate answers directly via a vanilla RAG pipeline in a zero-shot setting and evaluate their quality based on the standard answers.
As represented in Table~\ref{dqa_exp}, we test the generation performance of these LLMs in four conditions: 1) using no reference document (\textbf{NA}), 2) using top five documents retrieved by BM25, 3) using top five documents retrieved by BERT bi-encoders, 4) using reference documents provided by \textsf{Qilin} (\textbf{Oracle}).
Considering the answer quality at both syntactic and semantic levels, we use ROUGE-L and BERTScore (F1) as the evaluation metrics.
From the table, we have several observations.
Firstly, the generation performance increases in the order: NA < BM25 < BERT$_{bi}$ < Oracle, which is consistent with our expectation.
Through comparison, we find Qwen2.5 achieves the highest ROUGE-L score while Llama outperforms other models in terms of semantic matching, indicating that different LLMs may have different capability focuses.

\begin{table}[t]
\centering
\caption{Comparison of Retrieval-augmented generation (RAG) performance across various LLMs.}
\vspace{-0.2cm}
\label{dqa_exp}
\begin{small}
%\fontsize{7}{10.5}\selectfont
\begin{tabular}{l|l|cccc}
\Xhline{0.9pt}
\multicolumn{2}{c|}{\textbf{Generation}$\backslash$\textbf{Retrieval}} & \textbf{NA} & \textbf{BM25} & \textbf{BERT$_{bi}$} & \textbf{Oracle} \\
\hline
\multirow{4}{*}{ROUGE-L} & GPT3.5 & 0.332 & 0.388 & 0.390 & 0.424\\
& GPT4o-mini & 0.316 & 0.384 & 0.380 & 0.429\\
& Qwen2.5-72B & \textbf{0.342} & \textbf{0.396} & \textbf{0.396} & \textbf{0.436}\\
& Llama3.3-70B & 0.290 & 0.349 & 0.347 & 0.396 \\
& GLM4 & 0.318 & 0.363 & 0.366 & 0.401 \\
\hline
\multirow{4}{*}{BERTScore} & GPT3.5 & 0.708 & 0.725 & 0.722 & 0.748 \\
& GPT4o-mini & 0.710 & 0.729 & \textbf{0.728} & 0.754 \\
& Qwen2.5-72B & \textbf{0.714} & 0.725 & 0.724 & 0.752 \\
& Llama3.3-70B & 0.699 & \textbf{0.730} & 0.727 & \textbf{0.760} \\
& GLM4 & 0.701 & 0.715 & 0.713 & 0.739 \\
\Xhline{0.9pt}
\end{tabular}
\end{small}
\end{table}

\section{Conclusion}
In this work, we have presented a novel multimodal S\&R dataset \textsf{Qilin}.
Comprising APP-level sessions from 15,482 users, \textsf{Qilin} provides both textual and image content for heterogeneous results.
Besides, we have also collected abundant contextual signals such as query sources, multiple user feedback, and deep query answering (DQA) details to facilitate the investigation of various IR-related tasks.
Through comprehensive data analysis covering demographics, user engagement, result distribution, and query patterns, we present multi-faceted insights for enhancing S\&R systems.
To better instantiate its application, we conduct preliminary experiments in search, recommendation, and deep query answering on \textsf{Qilin}.
We believe these findings and insights will be valuable in developing more advanced multimodal retrieval systems. 

\newpage
\balance
\bibliographystyle{ACM-Reference-Format}
\bibliography{sample-base}
% \end{CJK}
\end{document}